\documentclass[preprint,aps,floats,subeqn,showpacs,amssymb]{revtex4}
\usepackage{epsfig}

\newcommand{\be}{\begin{equation}}
\newcommand{\ee}{\end{equation}}
\newcommand{\bea}{\begin{eqnarray}}
\newcommand{\eea}{\end{eqnarray}}
\newcommand{\bml}{\begin{mathletters}}
\newcommand{\eml}{\end{mathletters}}

\begin{document}

\tighten

\draft




\title{Inflating Branes inside hyper-spherically symmetric defects}
\renewcommand{\thefootnote}{\fnsymbol{footnote}}
\author{Y. Brihaye\footnote{yves.brihaye@umh.ac.be} , T. Delsate }
\affiliation{Faculte des Sciences, Universite de Mons-Hainaut, 7000 Mons, Belgium}

\date{\today}
\setlength{\footnotesep}{0.5\footnotesep}

\begin{abstract}
Static and inflating
branes residing at the center of a
hyper-spherical symmetric defect are considered in
$4+n$-dimensions with a non zero  bulk cosmological constant. 
Several vacuum solutions
can be constructed explicitely when the bulk cosmological constant and cosmological
constant on the brane are non zero. 
We consider as hyper-spherical symmetric defect
a global monopole for generic values of $n$
and a local monopole for $n=3$. We construct
new solutions which are regular or periodic in the
bulk radial coordinate.
\end{abstract}

 \maketitle
\section{Introduction}
The basic idea of brane world models \cite{dvali} is that our universe
is represented by a (3+1)-dimensional subspace (a three-brane)
embedded in a higher
dimensional space-time (the bulk). While string theory has 
compact extra dimensions
which are typically of the order of the Planck scale, brane world
models can have large or even infinite extra dimensions \cite{rs}. 
The reason is that
all matter fields are assumed to be confined to the brane, while
only gravity lives in all dimensions. Since Newton's law, however,
is well tested down to the mm-scale, gravity has to be localised
``well enough''
to the brane. 

One possibility discussed extensively in recent years \cite{shapo} 
assumes a topological defect to reside in the bulk and the brane
to be localised at the center of this defect. In that case
the cosmological constant in the bulk $\Lambda_{4+n} \neq 0$ and
the localisation of gravity on the brane is achieved  
by a fine tuning of $\Lambda_{4+n}$
with the other coupling constants appearing in the theory.
Similar results have been developed recently \cite{demello}
without a cosmological constant
but including both a global and a local monopole. For large
direct interaction of the two sectors of the theory, the global
monopole disappears from the system and ``leaves behind'' a negative
cosmological constant.
In all the above mentioned cases, the brane is a Minkowski brane.

In \cite{cho,chovil,japanese} brane worlds have been considered 
from another point of view: the bulk cosmological constant is zero,
while the branes are inflating, i.e. possess a physical cosmological
constant. This is of interest since experimental data points to the fact
that our universe contains a (small) cosmological constant.

In this paper, we extend the ideas of 
\cite{cho,chovil,japanese} by allowing the bulk
cosmological constant to be non-zero.
We thus have two independant cosmological 
constants, one in the bulk and one on the (3+1)-dimensional brane.
The underlying field theory describing the topological defect is considered
to be the Goldstone model (a global monopole) for $n$ extra dimensions
and  the SO(3)
gauge-Higgs theory describing non-abelian monopoles in the case of three
extra dimensions. We find that the existing solutions of 
\cite{cho,chovil,japanese} are smoothly deformed by the
bulk cosmological constant. In addition, we show numerically that
new types of regular solutions exist. 
These solutions have periodic matter and metric functions
and require a fine tuning of the two cosmological
constants and, to our knowledge, provide a new class of regular space-times
where a compactification of the extra dimensions occurs naturally.

The model and the equations are presented is Section II. 
Explicit vacuum solutions
are considered in Section III.
Melvin-type solutions are discussed in Section IV.
Section V and VI, respectively, describes the solutions when
a global monopole and local monopole is present in the bulk.
We show that regular solutions, periodic with respect to the radial
coordinate associated to the bulk exist.
This contrasts with the solutions obtained in \cite{bdh} where
theories involving two extra-dimensions are investigated and an
Abelian-Higgs model leads to vortex located in the bulk.
In this theory, solutions  become periodic only
far enough from the vortex core where the matter fields have reached
their vacuum values.
\section{Action principle for an $n$-dimensional brane}
\par
The $(4+n)$-dimensional action reads:
\be
S = \frac{1}{2\kappa^2}\int \sqrt{-g}
(R - 2\kappa^2\Lambda_{n+4} )d^{4+n}x  + S_{top}
\ee
where $S_{top}$ is the action  describing the topological
defect residing in the bulk, it depends on the specific model
studied and will be given in explicit form
later in the paper. $\kappa^2 = \frac{1}{M_*^{2+n}}$ with $M_*$ the
 $4+n$-dimensional Planck mass and $\Lambda_{n+4}$ denotes the bulk cosmological
constant.

The general form of the non-factorisable
metric that we consider in this paper is given by
\be
ds^2 = M(r)^2 ds_4^2 + dr^2 + L(r)^2d\Omega_{n-1}^2
\ee
where $ds_4^2$ describes the metric of the brane. The transverse space has rotational
invariance, $r$ describes the bulk radial coordinate, while $d\Omega_{n-1}$ 
denotes the line element associated with 
the $n-1$ angles $\theta_i$, $i=1,..,n-1$ of the transverse space.

The corresponding Einstein equations read \cite{cho}:
\bea
G_\mu^\mu &=& -\frac{1}{4}\frac{R^{(4)}}{M^2} + 3\frac{M''}{M}
+ 3\frac{M'^2}{M^2} + 3(n-1)\frac{M'L'}{ML} \nonumber \\
&+& (n-1)\frac{L''}{L} + \frac{(n-2)(n-1)}{2}\left(\frac{L'^2}{L^2}
- \frac{1}{L^2}\right) = \beta (T_\mu^\mu- \gamma) \ \ \ , \ \ \
\mu = 0, \dots , 3
\eea
\be
G_r^r = -\frac{1}{2}\frac{R^{(4)}}{M^2} + 6\frac{M'^2}{M^2}
+ 4(n-1)\frac{M'L'}{ML} + \frac{(n-2)(n-1)}{2}\left(\frac{L'^2}{L^2}
- \frac{1}{L^2}\right) = \beta (T_r^r- \gamma)
\ee
\bea
G_{\theta_i}^{\theta_i} &=& -\frac{1}{2}\frac{R^{(4)}}{M^2} + 4\frac{M''}{M}
+ 6\frac{M'^2}{M^2} + 4(n-2)\frac{M'L'}{ML} + (n-2)\frac{L''}{L} \nonumber \\
&+& \frac{(n-2)(n-3)}{2}\left(\frac{L'^2}{L^2} - \frac{1}{L^2}\right)
= \beta (T_{\theta_i}^{\theta_i} - \gamma)
\ \ \ , \ \ \ i = 1, \dots , n-1
\eea
where $\Lambda_{n+4} \equiv \gamma$ and $\kappa^2 \equiv \beta$.
The energy momentum tensor is given by
\begin{equation}
T_{MN}=-\frac{2}{\sqrt{-g}} \frac{\delta S_{top}}{\delta g^{MN}} \ , \ \ M,N=0,..,n+4
\end{equation}
and $R^{(4)}$ denotes the Ricci scalar corresponding to 
the metric on the brane, which we
choose to be
  \be
   ds_4^2 = - dt^2 + e^{2 H t} ( (dx^1)^2 + (dx^2)^2 + (dx^3)^2)
   \ee
such that $R^{(4)}= 12 H^2$ with $H=\sqrt{\frac{\Lambda_{phys}}{3}}$, where
$\Lambda_{phys}$ denotes the physical cosmological constant on the brane.


\section{Vacuum solutions}
Vacuum solutions correspond to $T_M^N = 0$.
According to the sign of the
$4+n$-dimensional cosmological constant they have
positive, null or negative
curvature. The result is very similar to the 4-dimensional case.
In the present case we emphasize a more general situation where
4-dimensional slices of the space have an (anti)-de Sitter geometry.
They can induce angular deficits in the $n$-dimensional subspace.
Because the equations do not explicitely depend on the radial variable $r$
the solutions given below can be arbitrality translated in $r$.

\subsection{$\Lambda_{n+4}  > 0$}
The Einstein equations above possess explicit solutions
in terms of trigonometric functions where $L$, $M$ have the same period.
The solution reads:
\be
\label{trigo}
M(r) = \frac{H}{\omega} \sqrt{\frac{3}{n+2}}  \sin \omega r \ \ , \ \
L(r) =  \frac{1}{\omega} \sqrt{\frac{n-2}{n+2}}   \sin \omega r \ \ , \ \
\omega^2 =  \frac{2\beta\gamma}{12+(n-1)(n+6)}
\ee

 Note that this solution depends crucially on the
 two cosmological constants and that the solution exists only for $n>2$.
 Indeed, in the $6$-dimensional case where $n=2$, the equation for $M$
 decouples \cite{bdh} and leads to $L \propto M'$, 
 which is not compatible with the solution above where the metric
 functions $M$, $L$ are characterized by the same phase.
 In the case $n=1$, the function $L(r)$ is not defined.

  Note that the solution is not regular since for $r= k \pi/\omega$ 
  ($k$ integer)
  we have $M(r)=L(r)=0$ and the Ricci scalar
  \be
 \label{ricci}
     R = -12 \frac{H^2}{M^2} + 12 \frac{M'^2}{M^2}     
     + 8 \frac{M''}{M}  + 8(n-1) \frac{M'L'}{M L}  + 2(n-1) \frac{L''}{L}
     + (n-1)(n-2) \left( \frac{L'^2}{L^2}-\frac{1}{L^2}\right)
  \ee
  is obvioulsly  singular for these values of $r$.

  Let us finally emphasize that the warp
  factor $M$ is directly proportional to the parameter $H$, both are
  related to the brane.

The solution has a natural geometric interpretation~:
they describe the surface of an $n+1$ dimensional sphere in the
extra dimensions and presents
an  angular deficit
relative to the angles $\theta_i$.
 More precisely, with the line element in the extra dimensions
\be
ds_n^2 = dr^2 + \frac{n-2}{n+2}\frac{1}{\omega^2}\sin^2(\omega
r)d\Omega_{n-1}^2
\ee
and setting  $\Theta = \omega r$, we find
\be
ds_n = \frac{1}{\omega^2}\left(d\Theta^2 +
\frac{n-2}{n+2}\sin^2(\Theta)d\Omega_{n-1}^2\right)  \ .
\ee
The solid angle corresponding to a fixed value of $\Theta$
presents an angular deficit
\be
\sin(\Theta)\int \sqrt{\frac{n-2}{n+2}}d\Omega_{n-1}
\ee
although $\sin(\Theta)\int d\Omega_{n-1}$ is expected.
It represents indeed a deficit since
\be
\frac{n-2}{n+2} <1 \ \ , \ \ \forall \ n \ .
\ee
In addition, the radius of compactification is found to be
the parameter $1/ \omega$ defined in (\ref{trigo}).





\subsection{$\Lambda_{4+n}=0$}
In the case of vanishing bulk cosmological constant, the solutions
turn out to be linear functions or $r$, given by
\be
M(r) = \sqrt{\frac{3H^2}{n+2}} \ r \ \ , \ \   
L(r) = \sqrt{\frac{n-2}{n+2}} \ r \ .
\ee
This solution does not fulfill the regularity condition at $r=0$
and is singular at the origin (e.g. the scalar curvature is singular).
Moreover, this solution describes a flat bulk with an angular deficit
in the extra dimensions.

\subsection{$\Lambda_{n+4} < 0$}
Finally, the case of a negative bulk cosmological constant we find:
\be
    M(r) = \frac{H}{\tilde \omega} \sqrt{\frac{3}{n+2}}
    \sinh{\tilde \omega r} \ \ , \ \
    L(r) = \frac{1}{\tilde \omega} \sqrt{\frac{n-2}{n+2}} 
    \sinh{\tilde \omega r} \ \ , \ \
     \tilde \omega^2 \equiv \frac{- 2 \beta \gamma}{12+(n-1)(n+6)}
\ee
This solution can also be obtained by a suitable analytic continuation
of  (\ref{trigo}).

\section{Melvin-type Universe in $n$ dimensions}
In this section, we look for the higher dimensional
analogues of the well know
4-dimensional Melvin space-time \cite{melvin}, i.e. 
we look for solutions of the form:
\be
M(r) = M_0(r-r_0)^\mu
\label{melm} \ \ , \ \
L(r) = L_0(r-r_0)^\lambda
\label{mell}
\ee
where $M_0$, $L_0$, $r_0$, $\mu$, $\lambda$ are constants to be determined.\\
In the following, we will distinguish the cases where 
solutions develop the above behaviour
for $r \to \infty$  (asymptotic
solution) and the case where the solutions develop a singularity 
in the neighbourhood of $r=r_0$ (see \cite{cho} for the $n=3$ case), respectively.
\subsection{Asymptotic solution}
Because of the occurence of non-homogeneous terms 
(e.g. $1/L^2$ and $\Lambda_{n+4}/M^2$) in the Einstein
equations, power-like solutions of the form above cannot be exact 
for generic values of $H$, $\Lambda_{n+4}$, $n$.
However, we will see that they exist
as asymptotic solutions and  appear as ``critical''
solutions in
the case of global and local monopoles (see Section VI).
Let us for a moment neglect the non-homogeneous terms in the
Einstein equations. Then inserting the power law decay above leads
to the following conditions for the exponants $\mu$, $\lambda$:

\be
3\mu(\mu-1) + 3\mu^2 + 3(n-1)\mu\lambda + (n-1)\lambda(\lambda-1)
+ \frac{(n-1)(n-2)}{2}\lambda^2 = 0
\ee
\be
6\mu^2 + 4(n-1)\mu\lambda + \frac{(n-1)(n-2)}{2}\lambda^2 = 0
\ee
\be
4\mu(\mu-1) + 6\mu^2 + 4(n-2)\mu\lambda + (n-2)\lambda(\lambda-1)
+ \frac{(n-2)(n-3)}{2}\lambda^2 = 0  \ .
\ee
The solutions then read:
\be
\label{melvin}
   \mu = \frac{2  \pm \sqrt{(n+2)(n-1)}}{2(n+3)} \ \ \ , \ \ \
   \lambda = \frac{(n-1)\mp 2 \sqrt{(n+2)(n-1)}}{(n-1)(n+3)}
\ee
Note, however, that with these exponents,
it is not justified to neglect the inhomogeneous
terms $\propto 1/L^2$ and $\propto 1/M^2$ except is the particular
case of the critical solutions (see Section V).

\subsection{Singular solutions}
If we want to interpret the functions (\ref{mell}),(\ref{melvin}) 
as the dominant
terms of a solution of the vacuum Einstein equations which are 
singular in the limit $r \to r_0$, the exponents should be such that
 $\mu < 1$, $\lambda < 1$.
It turns out that these conditions are fulfilled for {\it both}
values of the sign $\pm$ in (\ref{melvin}).

The standard Kasner conditions
read:
\be
4\mu^2 + (n-1)\lambda^2 =  1  \ \ , \ \
4\mu + (n-1)\lambda = 1  \ .
\ee
\par
It turns out that only the {\it linear} relation is fulfilled.
This means that
\be
g^m_{n+4} = r^{4-2n} g^p_{n+4}
\ee
where $g^m_{n+4}$  is the determinant of the metric corresponding
to an $n$-dimensional Melvin space while
 $g^p_{n+4}$ represents the
determinant corresponding to the $n$-dimensional flat space.


\section{Global monopoles}

Global monopole solutions occurs in the Goldstone model
describing $n$ scalar fields in a field theory globally invariant under
the $O(n)$ transformations. The  symmetry is
spontaneously broken by a potential. In the present context, the
Goldstone model and the corresponding scalar fields are formulated
with respect to the $n$ extra dimensions~:
\be
S_{top}=\int \left((\partial_A\Phi)^\dagger(\partial^A\Phi) -
\frac{\alpha}{4}(\Phi^\dagger\Phi - v^2)^2 \right)d^7x
\ee
where the $n$ scalar fields $\Phi = (\phi^1,\ldots,\phi^n)$
form a fundamental representation of the group $O(n)$. $\alpha$ is the
self-coupling of the potential, $v$ the vacuum expectation value of the
scalar field.

We use the Ansatz:
\be
\phi^i = \phi(r)\xi^i/r  \ ,
\ee
where the $\xi^i$ denote the cartesian coordinates
representing the extra dimensions.
Correspondingly, the energy momentum tensor has only
diagonal components given by (\cite{japanese}):
\be
T_\mu^\mu = \phi'^2 + \frac{(n-1)\phi^2}{2L^2} +
\frac{\alpha}{4}(\phi^2 - v^2)^2
\ee
\be
T_r^r = -\phi'^2 + \frac{(n-1)\phi^2}{2L^2}
+ \frac{\alpha}{4}(\phi^2 - v^2)^2
\ee
\be
T_{\theta_i}^{\theta_i} = \phi'^2
 + \frac{(n-3)\phi^2}{2L^2} + \frac{\alpha}{4}(\phi^2 - v^2)^2
\ee
where $\phi' \equiv d \phi / dr$. The equation corresponding
to the Goldstone field reads:
\be
    \phi'' + (n-1)(\frac{L'}{L} + \frac{4}{n-1}\frac{M'}{M})\phi' 
    - \frac{1}{L^2} \phi
    = \alpha \phi(\phi^2-v^2)
\ee

The appropriate boundary conditions read:
\be
M(0)=1 \ \ , \ \ M'(0)=0 \ \ , \ \ L(0) = 0 \ \ , \ \ L'(0)=1
\ee
for the metric functions. In the case when the radial variable
can be extended to $r=\infty$, the usual boundary conditions 
for the scalar field function are
\be
\phi(0)=0 \ \ , \ \  \phi(\infty) = v  \ .
\ee
However, the presence of a cosmological constant can lead to
a cosmological horizon at $r=r_c$. In such case, an appropriate
boundary condition for $\phi(r_c)$ has to be imposed; this will
be discussed in due course.

The expressions of the energy momentum tensor
 contain terms of the form $1/L^2$
which also appear in the Einstein tensor.  If the gravitational constant
$\kappa$ is chosen such that
\be
\kappa v = \sqrt{n-2}  \ \ , \ \
\ee
the two inhomogeneous terms cancel. This value determines the so-called
``critical monopole''.  In this case, (and assuming $H=\Lambda_{n+4}=0$
in addition) the asymptotic Melvin solutions are just solutions of the Einstein
equations \cite{cho,japanese}.


\subsection{Sub-critical monopoles}
The case  $\kappa\eta < \sqrt{n-2}$
corresponds to the case of subcritical
monopoles \cite{japanese}. The vacuum solutions for which the
functions $M$, $L$ go asymptotically to infinity (i.e. corresponding
to $\Lambda_{n+4} \leq 0$)   are such that the term
 $\phi/L^2 \to 0$.
The metric function
$M(r)$ remains the same irrespectively of the presence of
a global monopole. The function  $L(r)$
must be renormalized according to
\be
L(r)_m = \frac{L(r)_v}{\sqrt{1-\frac{(\kappa v)^2}{n-2}}}
\ee
Here $L_v$ corresponds to the function of the vacuum solution
while $L_m$ corresponds to the solution in the presence of the monopole. \\

In the case $\Lambda_{n+4} > 0$ the arguments above do not apply 
because the terms $\phi / L^2$ cannot be neglected. However, the
profiles of the metric functions $M$, $L$ remain 
 very close
to those of the vacuum solution.
For larger values of $r$, the metric  become singular at some finite value
of $r$. The singularity is of Melvin type and is of the same
nature as in the case of local monopoles (see discussion below).


\subsection{Mirror symmetric solutions}
The coupled system of equations
possesses several symmetries, namely invariance under translations of
the radial variable $r$ and invariance under the
reflexions $r \to -r$ and $\phi \to -\phi$. These symmetries suggest that
solutions  which are invariant under suitable combinations
of the symmetries could exist . 
In the case of vacuum solutions, the solutions
corresponding to $\Lambda_{n+4} > 0$  possess such a symmetry.
The most natural combination of the symmetries suggests to look
in particular for solutions with the following properties
\be
      L(r_0-r) = L(r_0+r) \ \ , \ \
      M(r_0-r) = M(r_0+r) \ \ , \ \
      \phi(r_0-r) = \epsilon \phi(r_0+r)  \ \ , \ \ \epsilon = \pm 1
\ee
where the reflexion point $r_0$ depends on the various coupling constants.
The existence of solutions presenting one of the above symmetries
can be analysed by solving the equations supplemented by the following
boundary conditions at $r=r_0$~:
\be
 L'(r_0)=0 \ , \ M'(r_0)=0 \ , \ \phi(r_0) = 0 \ \ {\rm if} \ \epsilon=-1
 \ \ , \ \ \phi'(r_0) = 0 \ \ {\rm if} \ \epsilon=1  \ .
\ee
These conditions complete the ones given above  for $r=0$ and allow for
a numerical study of the equations (no explicit solution was found for
$\phi \neq 0$).
Our numerical analysis suggests that (i)
solutions corresponding to $\phi(r_0-r) = \phi(r_0+r)$ do not seem to
 exist. In fact we were able to construct numerically such configurations
 but they do not persist when increasing the accurancy, 
(ii) solutions obeying    $\phi(r_0-r) = - \phi(r_0+r)$  do indeed
exist for peculiar values of the coupling constants.

The existence of these ``odd'' solutions can be related
to the fact that,  in the neighbourhood of the symmetric point
$r_0$, the Goldstone field equation can be put into the form
\be
     \phi ''- \alpha \phi^3
     + (\alpha v^2 - \frac{1}{L(r_0)^2})\phi  \sim 0
\ee
where we used the fact that $M'(r_0)=L'(r_0)=0$.
This simplified version of the
Goldstone equation
is identical to the
kink equation provided
$L(r_0) v > 1$ which turns out to hold true in the
region of parameters that we have explored.
Kink-like solutions can therefore be expected. 

Two solutions
of this type  are presented in Figs. 1 and 2. 
These solutions are similar to
 the so called ``Bag of Gold''- solutions discovered in the
4-dimensional Einstein-Yang-Mills equations with a positive cosmological
constant \cite{lavre}. Similar phenomena in 4-dimensional space-time
were observed in \cite{bfm} and more recently in \cite{fr}. However,
to our knowledge, it is the first time that compact solutions  
relative to the spatial extra dimensions are constructed.

It has to be stressed that these type of solutions exist only for
peculiar values of the cosmological constants  $\Lambda_{n+4}$, $H$ once
the coupling constants  $\alpha$, $\beta$
are fixed. Setting $\alpha=0.2$, $\beta=0.3$, the relations between
$\Lambda_{n+4}$, $H$ and $r_c$ allowing for mirror symmetric solutions
are given in Fig. 3. Our results suggest that several branches
of solutions could exist. We were able to find two non trivially different
branches, see Fig. 3.
The construction of the second branch is however
involved numerically.

It is clear that these solutions can be extended symetrically
for  $r \in [r_c,2 r_c]$ and further for $r \in [2 r_c, 4 r_c]$.
This leads naturally to periodic solutions on $[0, 4r_c]$. The corresponding
space-times correspond to hyperspheres with angular deficit.

\section{Local monopole for $n=3$}
It is well know that monopoles exist in some spontaneously broken
non-abelian gauge theories. The simplest case is the Georgi-Glashow
model with gauge group $SO(3)$ and a Higgs triplet. In the present
context, the action density reads
\be
S_{top}= \int \left((D_A\Phi^a)(D^A\Phi^a) - \frac{1}{4} F_{AB}^a F^{a,AB}
-\frac{\alpha}{4 e^2}(\Phi^a \Phi^a - v^2)^2\right)  d^7x
\ee
where $A,B=0,1,\dots,6$, $a=1,2,3$ and using
the usual definitions for the covariant derivative and fields strenghts~:
\be
   D_M \Phi^a = \partial_M  \Phi^a + e \epsilon^{abc} W^b_M \Phi^c \ \ , \ \
   F^a_{MN} = \partial_M W^a_N - \partial_N W^a_M + e \epsilon^{abc} W^b_M W^c_N
\ee
Along with \cite{chovil}, we use a spherically symmetric ansatz for
the gauge and Higgs fields~:
\be
   W_{\mu}^a = 0 \ \ , \ \ W_i^a = (1-w(r)) \epsilon_{a \ (i-3)(j-3)}
   \frac{x^j}{e r^2}
\ee
\be
    \Phi^a =   \phi(r) \frac{x^{3+a}}{r} \ \ \ \ , \ \ a = 1,2,3
\ee
leading to a energy momentum tensor with non-vanishing components:
\bea
      T_{\mu}^{\mu} &=& \frac{(\phi')^2}{2} + \frac{\phi^2 w^2}{L^2}
      + \frac{1}{e^2 L^2}( \frac{(1-w^2)^2}{2 L^2} + (w')^2)
      + \frac{\alpha }{4}(\phi^2-v^2)^2
        \nonumber \\
      T_{r}^{r} &=&  - \frac{(\phi')^2}{2} + \frac{\phi^2 w^2}{L^2}
      + \frac{1}{e^2 L^2}( \frac{(1-w^2)^2}{2 L^2} - (w')^2)
      + \frac{\alpha}{4}(\phi^2-1)^2
        \nonumber  \\
      T_{\theta_i}^{\theta_i} &=&  \frac{(\phi')^2}{2}
           - \frac{(1-w^2)^2}{2 e^2 L^4}
          +  \frac{\alpha }{4}(\phi^2-v^2)^2  \ .
\eea
The classical equations can be obtained easily by substitution of the 
expressions above in (3)-(5).
In the absence of a bulk comological constant, this model was studied in
\cite{chovil}.
In this paper, the equations are solved with the usual boundary
conditions for the functions $w(r),\phi(r)$:
 $\phi(0)= 0$ , $w(0) = 1$ for regularity at the origin and
$\phi(r\to \infty)= v$ , $w(r \to \infty) = 0$
outside  the monopole core.

Here we will analyze the influence
of the bulk cosmological constant
on the solutions.  Again, the presence
of cosmological constant will force us to impose 
conditions for the matter fields at an intermediate
value $r=r_c$.
In the discussion and the figures the following rescaling
of the parameters: $\beta = e^2 v^2 \kappa^2$, 
$\gamma = \Lambda_{4+n}/e^2 v^2$,
 $H/ev \to H$ are used.

In general, the three types of vacuum solutions presented in Section IV
are deformed by the presence of the local monopole on the brane.
The principle deformation of the metric fields
for $\Lambda_{n+4}=0,\pm 0.01$
are sketched in Fig.4 with $H= 0.15$.
We will now discuss the different solutions more qualitatively.

\subsection{$\Lambda_{n+4} = 0$}
In this case, the solutions have been analyzed in details
in \cite{chovil}. Far from the monopole core, the metric
functions behave linearly:
\be
     M(x) = \sqrt{\frac{3}{5}} H r + m_0 + \frac{m_1}{r}\ \ \ , \ \ \
     L(x) = \sqrt{\frac{1}{5}} r + l_0 + \frac{l_1}{r}  \ .
\ee
These approach the corresponding vacuum solutions of Section 4 for $r\to 0$.

\subsection{$\Lambda_{n+4} < 0$}
For $\Lambda_{n+4} < 0$ our numerical analysis reveals that there
are regular solutions which obey asymptotically the form
\be
   M(r) = M_0 \sinh{\mu r} \ \ , \ \ L(r) = L_0 \sinh{\mu r} \ \ , \ \
   \mu = \sqrt{\frac{-\beta \gamma}{15}}    \ .
\ee
These solutions have thus exponential behaviour. 
Two comments are in order here:
\begin{itemize}
\item These solutions make sense only with the positive square root
of $\mu$. Indeed, the non-homogeneous terms $1/L^2$ and $1/L^4$
occuring in the equations can then be neglected asymptotically. 
With the negative
square root, these non-homogeneous terms cannot be neglected.
\item These solutions persist in the $\Lambda_{4+n} = 0$ limit, forming
another branch of solutions with respect to the one studied in \cite{shapo},
where the emphasis is set on warped solutions.
The profile of such a solution is presented in Fig.5 for
$H=0$ and $\Lambda_{n+4}=-0.01$. In particular, we can see the deviation
of the functions $M$, $L$  from their asymptotic behaviour.
\end{itemize}

\subsection{$\Lambda_{n+4} > 0$}
We studied the solution for $\Lambda_{n+4} >0$ and found qualitatively how
the vacuum trigonometric solutions are deformed by the matter
fields. This is illustrated in Fig.6. As expected, we observe that
the monopole regularizes the singular vacuum solutions such 
that the metric fields become regular on the brane $r=0$.
The functions $M$, $L$ reach their maximum value when the matter fields
have already reached their asymptotic values.
The mirror symmetry of the vacuum solutions is broken, in particular
the numerical analysis reveals that
the two functions $M$, $L$ do not reach their maximum exactly at
the same value of $r$.
It is likely that mirror symmetric solution exist in this case
as well for tuned values of the constants
$H$, $\Lambda_{n+4}$. Mirror symmetric
solutions will be discussed in the next section.

The challenging question is to understand
how these  solutions (regularized at the origin) behave in the
neighbourhood of the first period of the associated
periodic solution of the vacuum equation. This is illustrated in Fig. 7.
This clearly indicates that the solution develops a singularity
at some $r=r_c$, for $\gamma=0.07,\Lambda_{n+4}=0.005$
we have $r_c\approx 291.7$.
A detailed analysis of the numerical solution
for $r \propto r_c$ reveals that the singularity is of Melvin type
discussed in the previous section. For generic values of the
inflating parameters and positive bulk cosmological constant, the 
local monopole solutions present a singularity at some finite value
of the radial variable relative to the extra dimensions.
\subsection{Mirror symmetric solutions}
In the previous sections we reported solutions obtained for generic
values of the different coupling constants. However, it is likely
that for specific values of $H$, $\Lambda_{n+4}$
regular solutions exist. Mirror symmetric solutions,
 like the ones obtained in the context of global monopoles are good
 candidates. It turns out that also in the case of local monopoles,
 it is possible to tune $H$ and $\Lambda_{n+4}$ in such a way that
 a mirror symmetric solutions exist. 
 In order to construct such solution , we imposed the
 constraints  $\phi=0$, $w=0$ $M'=0$, $L'=0$ as  supplementary boundary
 conditions at some finite value $r=r_c$; then we tried to determine
 numerically if the values of $H, \gamma$ can be adjusted for all
 constaints to be fullfilled. It turns out to be possible.
 For fine tuned  values of the two cosmological constants, indeed, 
 our numerical integration of the equations 
 exhibits solutions such that the function $\phi(r)$ bends backwards 
 and reaches
 the value $\phi=0$ at  $r=r_c$ after developping
 a plateau where $\phi \sim 1$ for $0 < r < r_c$. At the same time
 the function $w(r)$ and the
 derivatives of the metric functions $L$ and $M$, $L'$, $M'$
approach zero for $r \to r_0$.

 An example of such a solution is given in Figs. 8 and 9.
 It corresponds to $r_c = 100$ which fixes $\gamma \approx -0.003$,
 $H \approx 0.187$. We constructed a branch of solutions for lower values
 of $r_c$ and obtained , e.g. for $r_c=50$ the values
 $\gamma \approx 0.018$, $H \approx 0.217$.
 The sign of the bulk cosmological constant $\Lambda_{4+n}$ 
 leading to these kind of solutions
 seems to be negative for $r_c > 82$ and
 positive for  $r_c < 82$. The value $H$ is positive
 and  does not vary significantly with $r_c$.

 The properly speaking mirror symmetric solutions
 can then be obtained by extending the numerical solution discussed
 above  by mirror symmetry for $r \in [0,2 r_c]$ ($M,L,w$ are symmetric
 under the reflexion $r \to 2 r_c-r$ while $\phi$ is antisymmetric).
 This is possible by using
 of the discrete symmetries of the equations as discussed 
 above. These mirror symmetric
 solutions can further be extended periodically
 on the interval $r\in [0, 2p r_c]$ for $p$ integer.
  The periodicity of the function $L(r)$ suggest that
 the geometry of these solutions in the extra dimension
 consists of the surface of a deformed 3-sphere. We can then
 interpret the radial coordinate $r$ as the "colatitude" and
 the solution looks like a magnet (with non-abelian fields)
 whose positive and negative poles lay respectively on the
 north ($r=0$) and south ($r=2r_c$) poles of the sphere.
 In the region of the equator ($r=r_c$) the Higgs field
 form a domain wall which separates these two poles.  This is
 illustated by fig. 10 where the Ricci scalar $R$ (see Eq. (\ref{ricci}))
 , the
 full energy density $T^0_0$ and the contribution 
to the energy of  the Yang-Mills
 fields $T^0_{0,ym}$ of the solution of Figs. 8,9
 are represented in the north-pole region and in the equator
 region.   In particular, we see that only the Yang-Mills energy
 is zero in the equator region, so that the energy content
 in this region is due to the Goldstone boson.

 \section{Conclusions}
 In this paper, we have reconsidered a brane world
model in $(4+n)$ dimensions. We have considered the cases
of a bulk without matter, a global monopole and a local monopole,
respectively.
 The difference with respect to existing literature is the inclusion
 of both a bulk cosmological constant and a physical cosmological
constant on the brane.
 For generic values of these constants, the existing solution get
 smoothly deformed and leads to a pattern similar to the case
 of standard cosmological
 solutions of 4-dimensional space-time,
 largely determined by  the sign of $\Lambda_{4+n}$.

 The presence of two constants allows for new types
 of solutions, namely solutions  which are
 mirror symmetric under the reflexion $r \to 2r_0-r$, where $r_0$
 is fune tuned with respect to the cosmologival constants $\gamma$
 and $\Lambda_{4+n}$.
 These mirror symmetric solution can eventually
 be extended into periodic solutions  in the radial coordinate
 associated with the  extra dimensions; this
 provides a natural compactification of the transverse space.
\\
\\
{\bf Acknowlegdement} Y.B. gratefully acknowledges
 discussions with
 A. Achucarro and
 B. Hartmann.\\
\\

\newpage
\begin{figure}
\centering
\epsfysize=17cm
\mbox{\epsffile{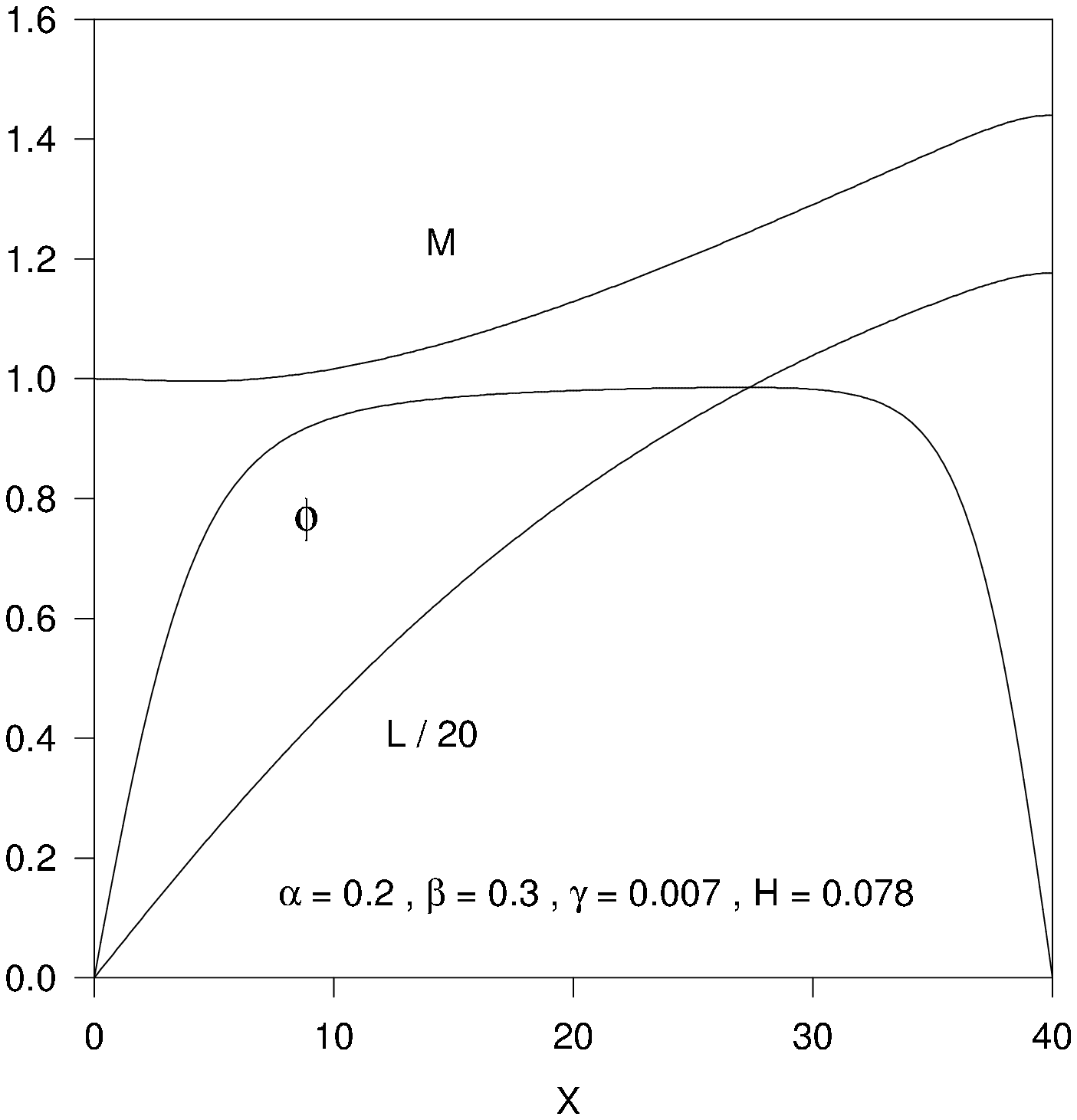}}
\caption{\label{fig1}
The profiles of the metric and scalar functions for the miror
symmetric solution corresponding to $x_c = 40$. This is a solution
on the first branch. }
\end{figure}
\begin{figure}
\centering
\epsfysize=17cm
\mbox{\epsffile{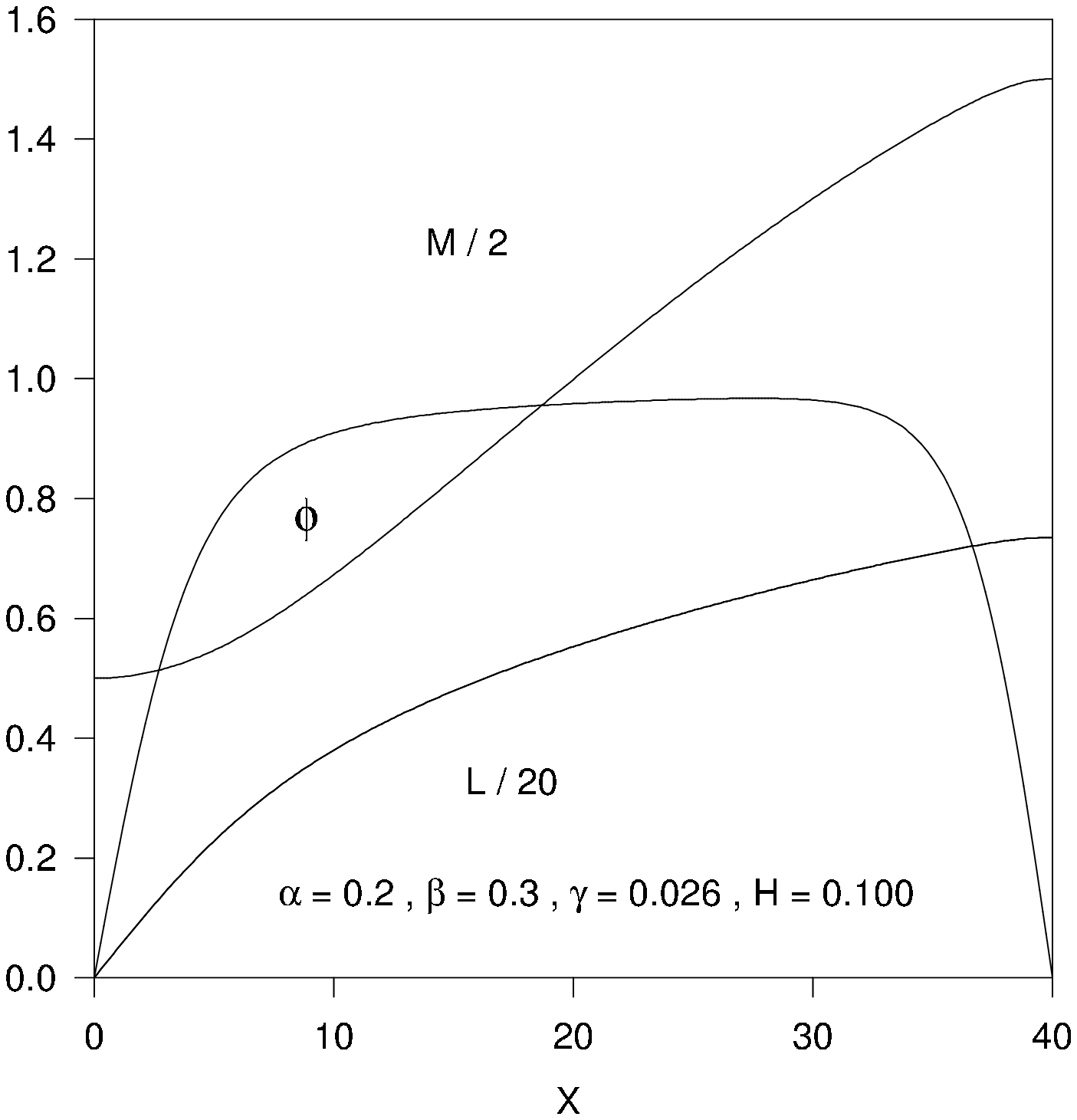}}
\caption{\label{fig2}
The profiles of the metric and scalar functions for the mirror
symmetric solution corresponding to $x_c = 40$. This is a solution
on the second branch. }
\end{figure}
\begin{figure}
\centering
\epsfysize=17cm
\mbox{\epsffile{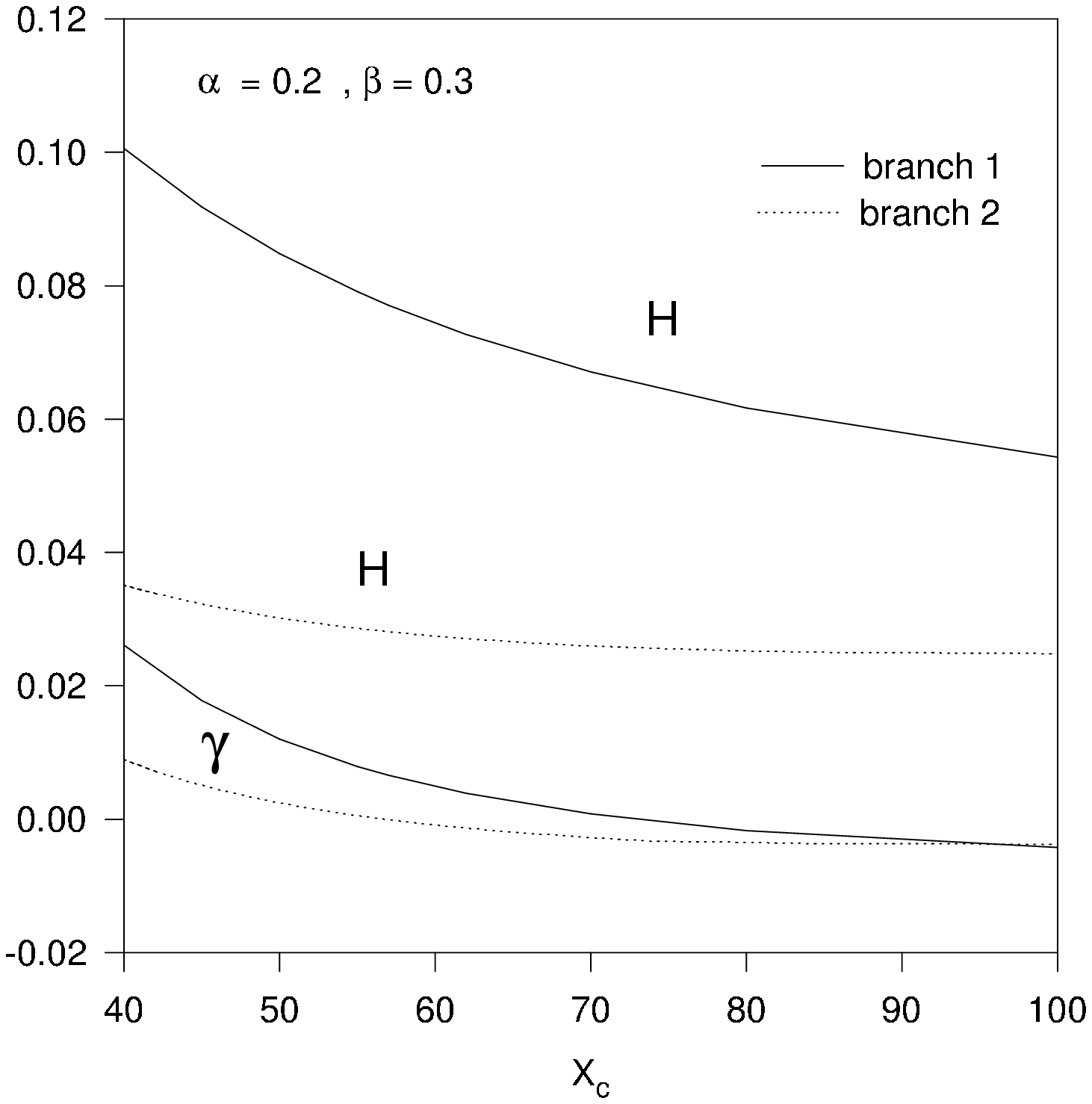}}
\caption{\label{fig3}
The relations between the parameters $\Lambda_{n+4}$, $H$, $r_0$ allowing mirror
symmetric solutions.}
\end{figure}

\begin{figure}
\centering
\epsfysize=17cm
\mbox{\epsffile{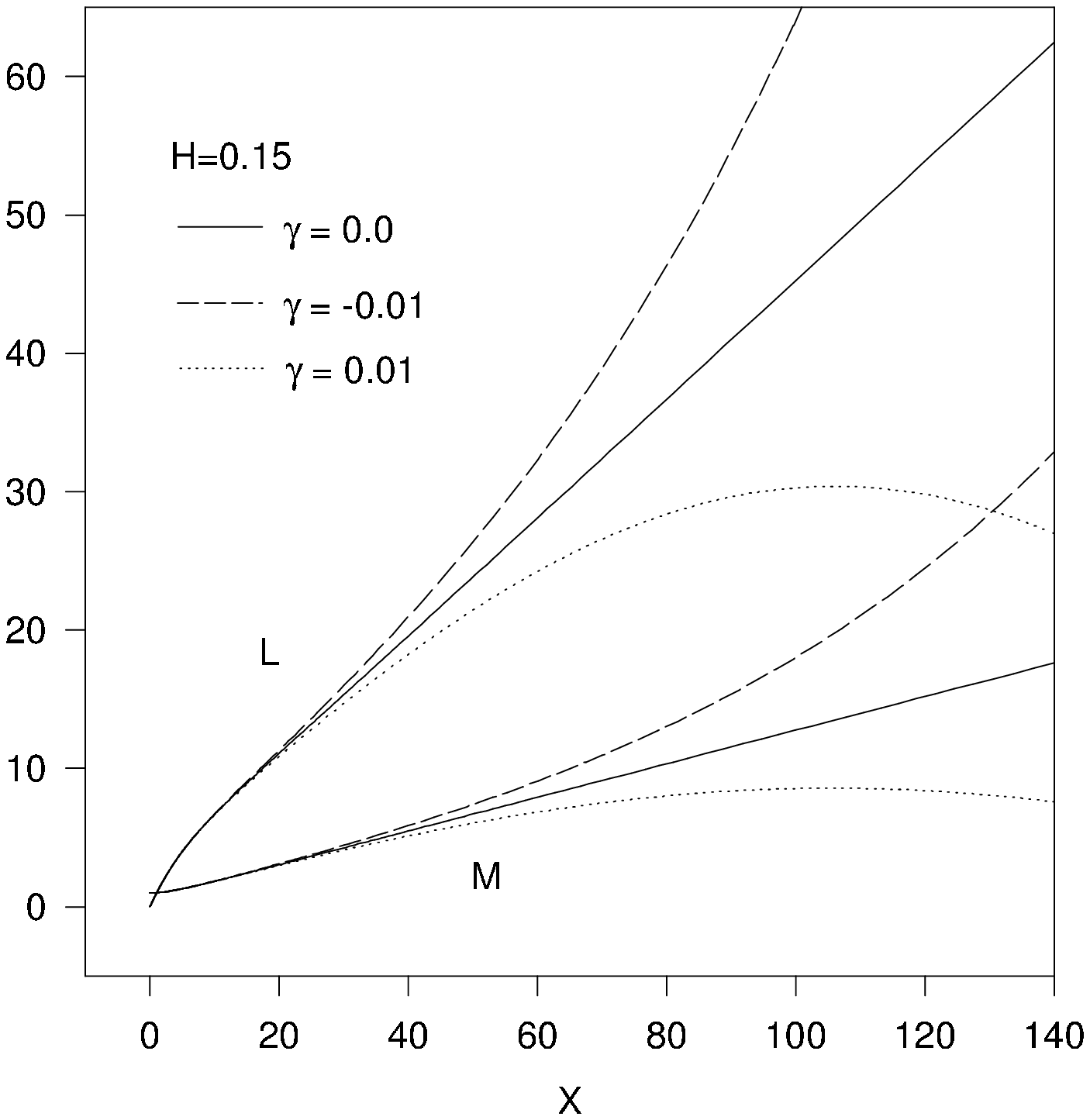}}
\caption{\label{fig4}
The profiles of the metric functions $L$, $M$ corresponding to an inflating
brane with $H=0.15$ regularized by a monopole are given for
$\Lambda_{n+4} = 0, \pm 0.01$.}
\end{figure}

\begin{figure}
\centering
\epsfysize=17cm
\mbox{\epsffile{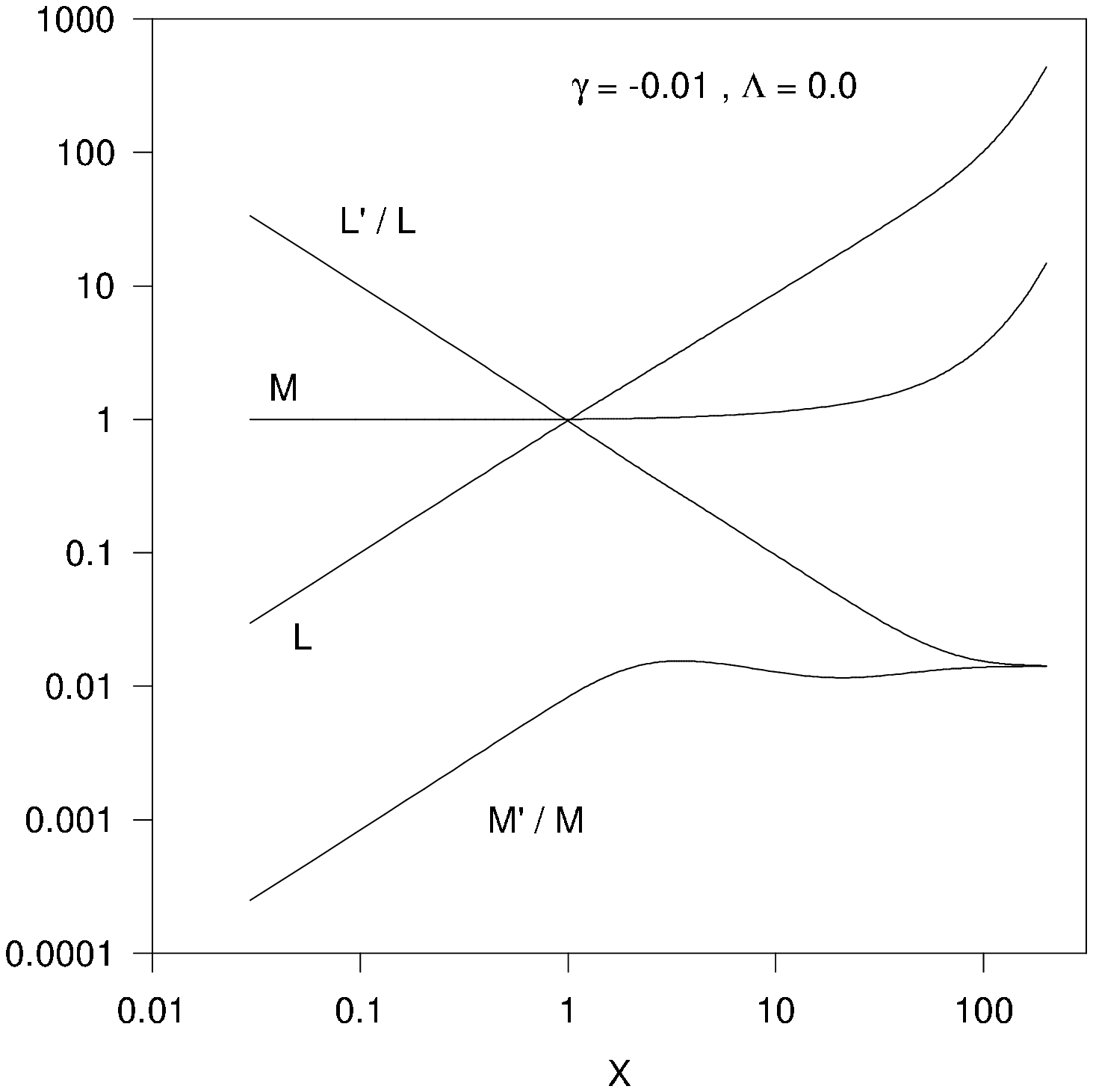}}
\caption{\label{fig5}
The profiles of the metric functions $L$, $M$ and the ratio $L'/L, M'/M$ 
corresponding to a static brane are shown for $\Lambda_{n+4} = -0.01$}
\end{figure}

 \begin{figure}
\centering
\epsfysize=17cm
\mbox{\epsffile{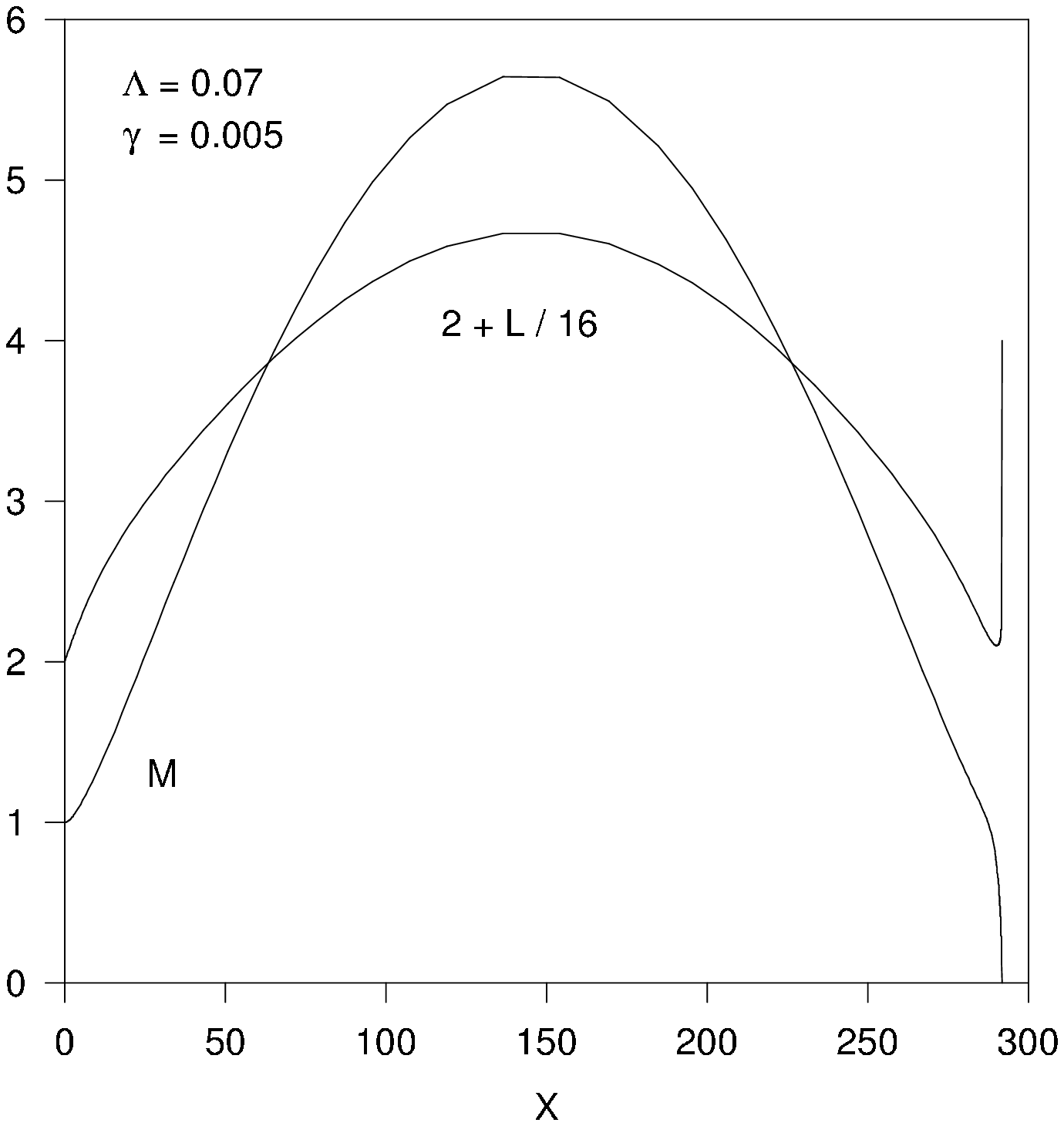}}
\caption{\label{fig6}
The profiles of the metric functions of an inflating brane solution with
$H=0.07$ and positive cosmological constant $\Lambda_{n+4} = 0.005$ 
are shown.}
\end{figure}

\begin{figure}
\centering
\epsfysize=17cm
\mbox{\epsffile{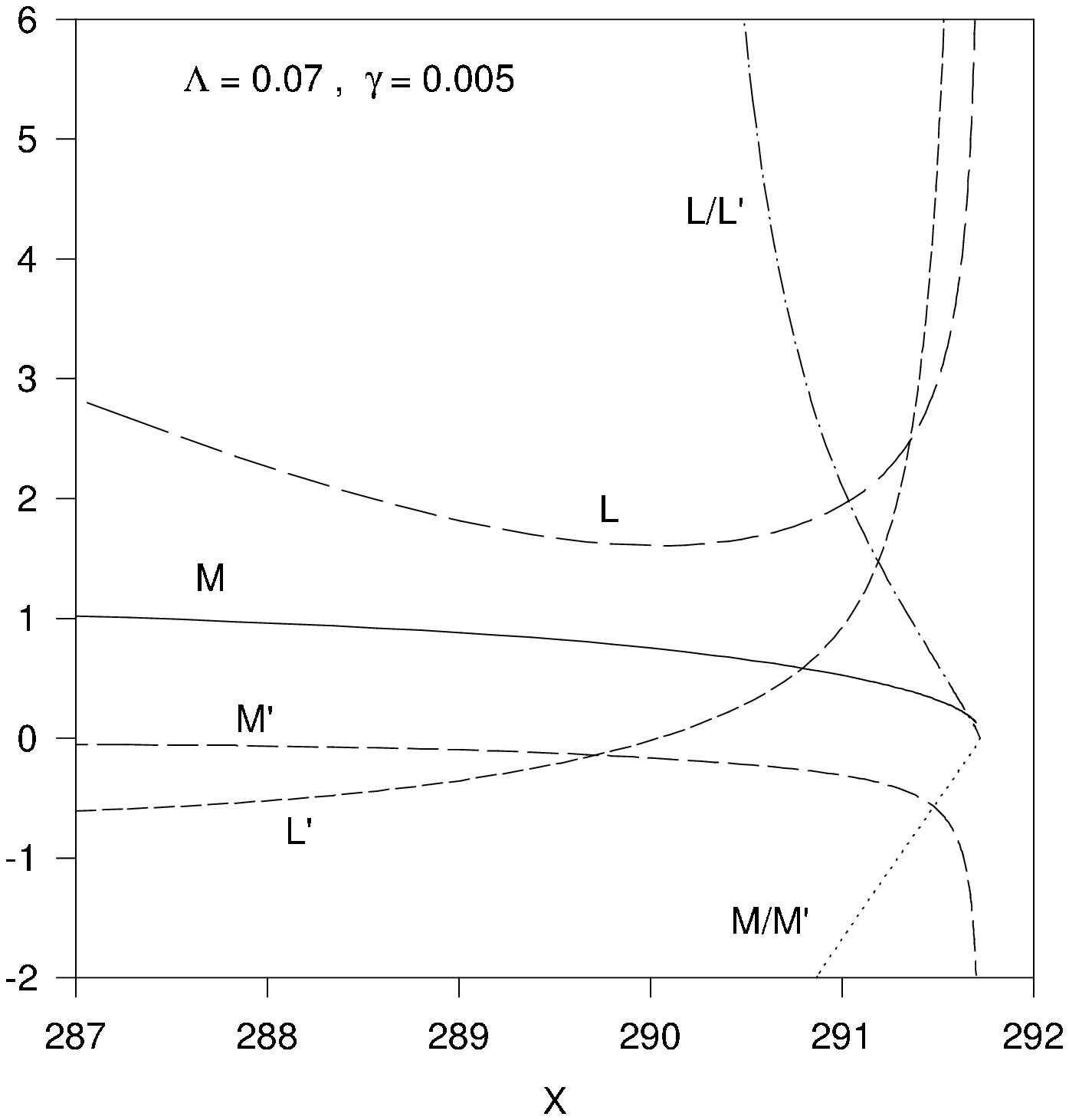}}
\caption{\label{fig7}
The details of Fig.6 in the region of the singular point $r\approx 291.7$.}
\end{figure}

 \begin{figure}
\centering
\epsfysize=17cm
\mbox{\epsffile{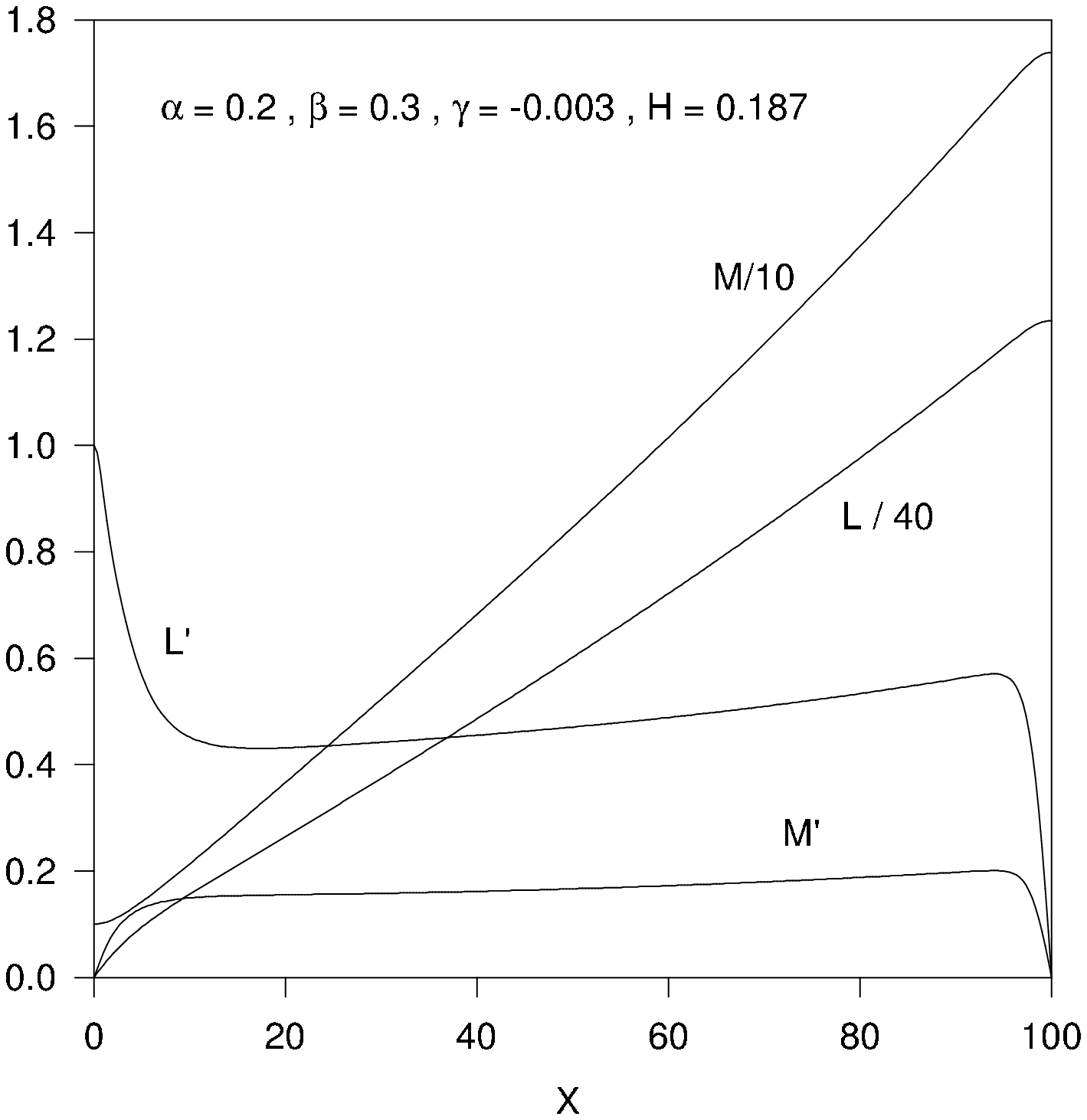}}
\caption{\label{fig8}
The profiles of the metric functions $M,L$ and their derivatives
for a periodic monopole solution.}
\end{figure}

\begin{figure}
\centering
\epsfysize=17cm
\mbox{\epsffile{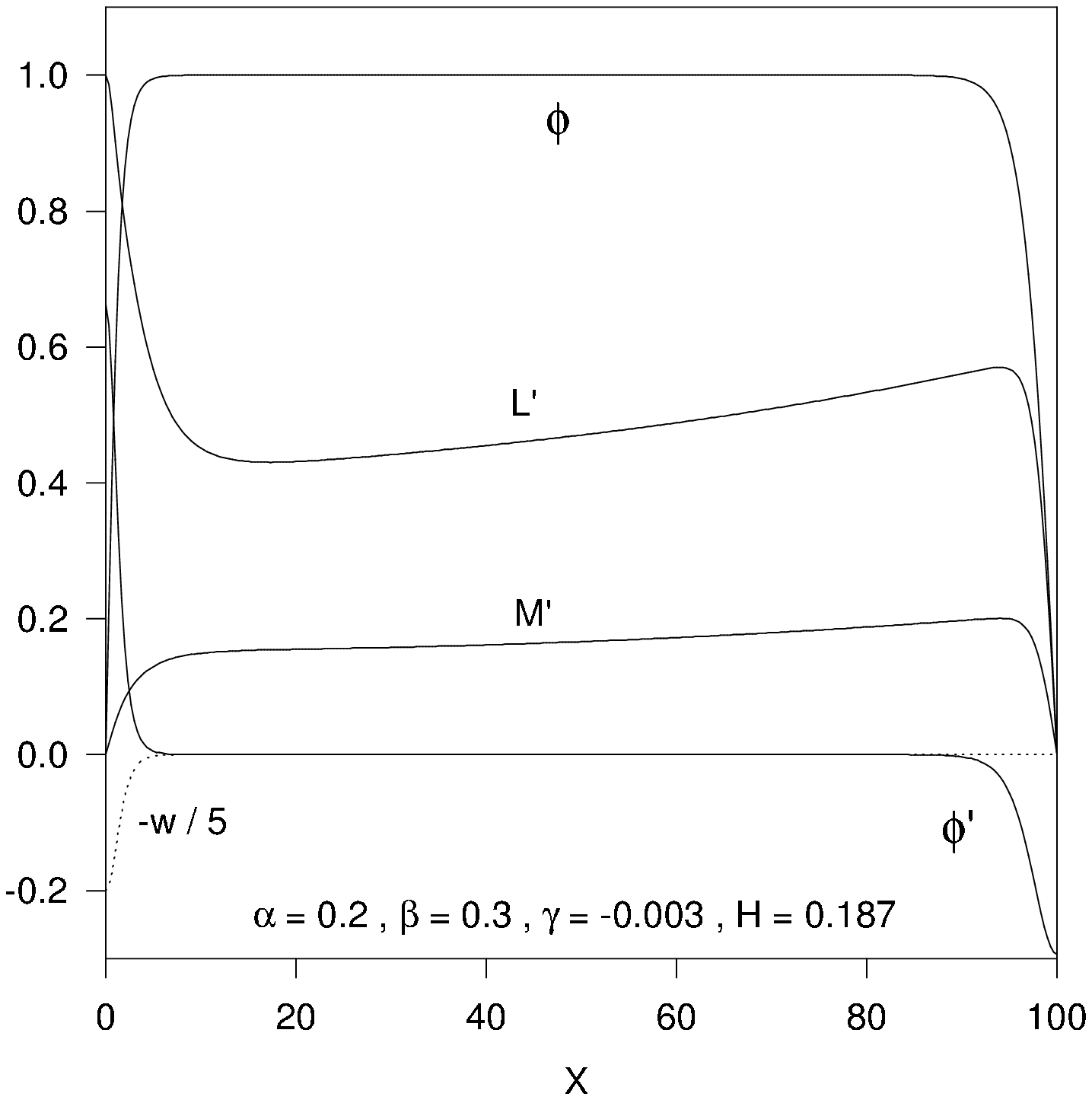}}
\caption{\label{fig9}
The profiles of the matter functions $w,\phi,\phi'$
and the derivatives $M',L'$ for a periodic monopole solution.}
\end{figure}

\begin{figure}
\centering
\epsfysize=17cm
\mbox{\epsffile{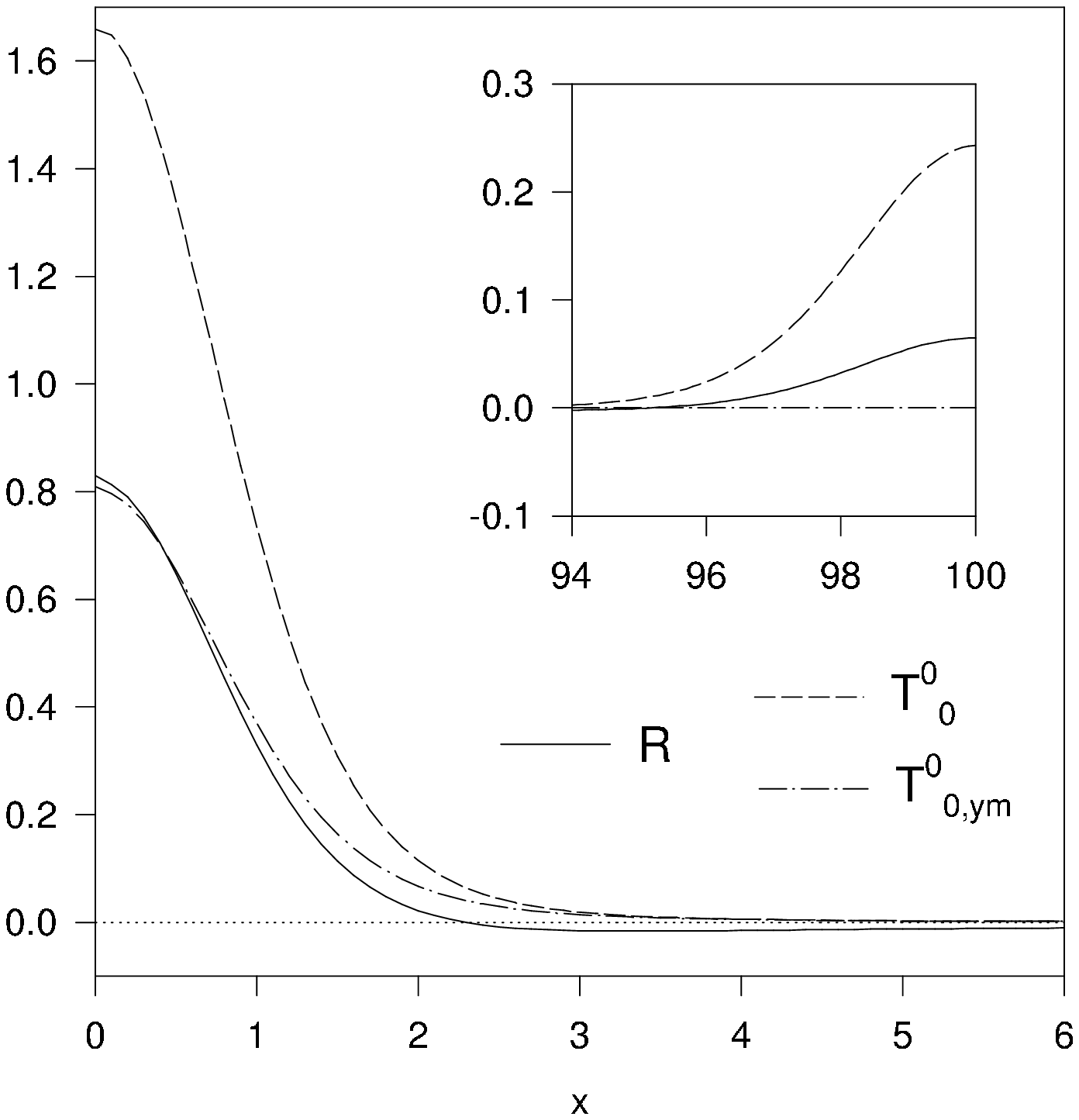}}
\caption{\label{fig10}
The Ricci scalar $R$, the energy momentum tensor $T^0_0$ and
the Yang-Mills contribution to the energy $T^0_{0,ym}$
corresponding to the periodic monopole of Figs. 8,9.}
\end{figure}

\end{document}